# Professional and Citizen Bibliometrics: Complementarities and ambivalences in the development and use of indicators. A state-of-the-art report.

Loet Leydesdorff,[a]* Paul Wouters,[b] and Lutz Bornmann[c]

(the authors contributed equally)

**Abstract**

Bibliometric indicators such as journal impact factors, $h$-indices, and total citation counts are algorithmic artifacts that can be used in research evaluation and management. These artifacts have no meaning by themselves, but receive their meaning from attributions in institutional practices. We distinguish four main stakeholders in these practices: (1) producers of bibliometric data and indicators; (2) bibliometricians who develop and test indicators; (3) research managers who apply the indicators; and (4) the scientists being evaluated with potentially competing career interests. These different positions may lead to different and sometimes conflicting perspectives on the meaning and value of the indicators. The indicators can thus be considered as boundary objects which are socially constructed in translations among these perspectives. This paper proposes an analytical clarification by listing an informed set of (sometimes unsolved) problems in bibliometrics which can also shed light on the tension between simple but invalid indicators that are widely used (e.g., the $h$-index) and more sophisticated indicators that are not used or cannot be used in evaluation practices because they are not transparent for users, cannot be calculated, or are difficult to interpret.

[a] *corresponding author;
Amsterdam School of Communication Research (ASCoR), University of Amsterdam, P.O. Box 15793, 1001 NG Amsterdam, The Netherlands; loet@leydesdorff.net.
[b] Centre for Science and Technology Studies CWTS, Leiden University, P.O. Box 905, 2300 AX Leiden, The Netherlands; p.f.wouters@cwts.leidenuniv.nl.
[c] Division for Science and Innovation Studies, Administrative Headquarters of the Max Planck Society, Hofgartenstr. 8, 80539 Munich, Germany; bornmann@gv.mpg.de ;

# 1. Introduction

In *Toward a Metric of Science: The Advent of Science Indicators* (Elkana, Lederberg, Merton, Thackray, & Zuckerman, 1978), the new field of science indicators and scientometrics was welcomed by a number of authors from the history and philosophy of science, the sociology of science (among them Robert K. Merton), and other fields. As the Preface states: "Despite our reservations and despite the obviously fledgling state of 'science indicator studies,' the conference was an intellectual success. Discussion was vigorous both inside and outside the formal sessions." The conference on which the volume was based, was organized in response to the first appearance of the *Science Indicators* of the US National Science Board in 1972, which in turn was made possible by the launch of the *Science Citation Index* in 1964 (de Solla Price, 1965; Garfield, 1979a).

The reception of scientometrics and scientometric indicators in the community of the sciences has remained ambivalent during the four decades since then. Science indicators are in demand as our economies have become increasingly knowledge-based and the sciences have become capital-intensive and organized on a large scale. In addition to input indicators for funding schemes (OECD, 1963, 1976), output indicators (such as publications and citations)[1] are nowadays abundantly used to inform research-policy and management decisions. It is still an open question, however, whether the emphasis on measurement and transparency in S&T policies and R&D management is affecting the research process intellectually or only the social forms in which research results are published and communicated (van den Daele & Weingart, 1975). Is the divide between science as a social institution (the "context of discovery") and as intellectually organized (the "context of justification") transformed by these changes in the research system? Gibbons *et al.* (1994), for example, proposed to consider a third "context of application" as a "transdisciplinary" framework encompassing the other contexts emerging since the ICT revolution. Dahler-Larsen (2011) characterizes the new regime as "the evaluation society."

---

[1] These indicators are sometimes named output indicators, performance indicators, scientometric indicators, or bibliometric indicators. We shall use the last term, which focuses on the textual dimension of the output.

In addition to these macro-level developments, research evaluation is increasingly anticipated in scientific research and scholarly practices of publishing as well as in the management of universities and research institutes, both intellectually (in terms of peer review of journals) and institutionally (in terms of grant competition). Thus, management is no longer external with respect to the shaping of research agendas, and scientometric indicators are used as a management instrument in these interventions. Responses from practicing scientists have varied from outright rejection of the application of performance indicators to an eager celebration of them as a means to open up the "old boys' networks" of peers; but this varies among disciplines and national contexts. The recent debate in the UK on the use of research metrics in the Research Excellence Framework provides an illustration of this huge variation within the scientific and scholarly communities (Wilsdon *et al.*, 2015). The variety can be explained by a number of factors, such as short-term interests, disciplinary traditions, the training and educational background of the researchers involved, and the extent to which researchers are familiar with quantitative methodologies (Hargens and Schuman, 1990).

Publication and citation scores have become ubiquitous instruments in hiring and promotion policies. Applicants and evaluees can respond by submitting and pointing at other possible scores, such as those based on Google Scholar (GS) or even in terms of the disparities between Web of Science (WoS, Thomson Reuters) and Scopus (Elsevier) as two alternative indicator sources (Mingers & Leydesdorff, 2015). In other words, it is not the intrinsic quality of publications but the schemes for measuring this quality that have become central to the discussion. Increasingly, the very notion of scientific quality makes sense only in the context of quality control and quality measurement systems. Moreover, these measurement systems can be commodified. Research universities in the United States and the United Kingdom, for example, use services such as Academic Analytics that provide customers with business intelligence data and solutions at the level of the individual faculty. This information is often not accessible to evaluees and third parties, so it cannot be controlled for possible sources of bias or technical error.

We argue that the ambivalences around the use of bibliometric indicators are not accidental but inherent to evaluation practices (Rushforth and de Rijcke 2015). In a recent attempt to specify the "rules of the game" practices for research metrics, Hicks, Wouters, Waltman, de Rijcke, and Rafols (2015) proposed using "ten principles to guide research evaluation," but also warn against "morphing the instrument into the goal" (p. 431). We argue that "best practices" are based on compromises, but tend to conceal the underlying analytical assumptions and epistemic differences. From this perspective, bibliometric indicators can be considered as "boundary objects" that have different implications in different contexts (Gieryn, 1983; Star & Griesemer,

1989). Four main groups of actors can be distinguished, each developing its own perspective on indicators:

1. *Producers*: The community of indicator producers in which industries (such as Thomson Reuters and Elsevier) collaborate and exchange roles with small enterprises (e.g., ScienceMetrix in Montreal; VantagePoint in Atlanta) and dedicated university centers (e.g., the Expertise Center ECOOM in Leuven; the Center for Science and Technology Studies CWTS in Leiden). The orientation of the producers is toward the development and sales of bibliometric products and advice;
2. *Bibliometricians*: An intellectual community of information scientists (specialized in "bibliometrics") in which the pros and cons of indicators are discussed, and refinements are proposed and tested. The context of bibliometricians is theoretically and empirically driven research on bibliometric questions, sometimes in response to, related to, or even in combination with commercial services;
3. *Managers*: Research management periodically and routinely orders bibliometric assessments from the (quasi-)industrial centers of production. The context of these managers is the competition for resources among research institutes and groups. Examples of the use of bibliometrics by managers are provided by Kosten (2016);
4. *Scientists*: The scientists under study who can be the subject of numerous evaluations. Nowadays, many of them keep track of their citation records and the value of their performance indicators such as the *h*-index. Practicing scientists are usually not interested in bibliometric indicators *per se*, but driven by the necessity to assess and compare research performance quantitatively in the competition for reputation and resources.

The public discourse about research evaluation and performance indicators is mainly shaped by translation processes in the communications among these four groups. The translations may move the discussion from a defensive one (e.g., "one cannot use these indicators in the humanities") to a specification of the conditions under which assessments can be accepted as valid, and the purposes for which indicators might legitimately be used. Under what conditions—that is, on the basis of which answers to questions—is the use of certain types of indicators justifiable in practice? However, from the perspective of developing bibliometrics as a specialty, one can also ask: under what conditions is the use of specific indicators conducive to the creation of new knowledge or innovative developments? If one instead zooms in on the career structures in science, the question might be: what indicators make the work of individual researchers visible or invisible, and what kind of stimulus does this give to individual scholars?

In the following, we list a number of important ambivalences around the use of bibliometric indicators in research evaluation. The examples are categorized using two major topics: the data and the indicators used in scientometrics. In relation to these topics, tensions among the four groups acting in various contexts can be specified. The main tension can be expected between bibliometric assessments that can be used by management with potentially insufficient transparency, *versus* the evaluees who may wish to use qualitative schemes for the evaluation. Evaluees may feel unfairly treated when they can show what they consider as "errors" or "erroneous assumptions" in the evaluations (e.g., Spaan, 2010). However, evaluation is necessarily based on assumptions. These may seem justified from one perspective, whereas they may appear erroneous from a different one. In other words, divergent evaluations are *always* possible. A clarification of some of the technical assumptions and possible sources of error may contribute to a more informed discussion about the limitations of research evaluation.

**2. Ambivalences around the data**

Despite the considerable efforts of the database providers to deliver high-quality and disambiguated data, a number of ambivalences with respect to bibliometric data have remained.

*2.1. Sources of bibliometric data*

WoS and Scopus are the two established literature databases for bibliometric studies. Both databases are transparent in the specification of the publications and cited references that are routinely included. Producers (group 1) and bibliometricians (group 2) claim that both databases can be used legitimately for evaluative purposes in the natural and life sciences, but may be problematic in many social sciences and humanities (Larivière, Archambault, Gingras, & Vignola-Gagné, 2006; Nederhof, 2006). The publication output in these latter domains would not be sufficiently covered. However, both providers make systematic efforts to cover more literature (notably books) in the social sciences and humanities.

With the advent of Google Scholar (GS) in 2004, the coverage problem may have seemed to be solved for all disciplines. GS has become increasingly popular among managers and scientists (groups 3 and 4). Important reasons for using GS are that it is freely available and comprehensive. Conveniently, the citation scores retrieved from GS are always higher than those from WoS and Scopus because the coverage is larger by an order of magnitude. However, it has remained largely unknown on which set of documents the statistics provided by GS are based. Materials other than research papers (e.g., presentations, reports, theses) are also considered for inclusion (Jacso, 2012b).

GS is regularly used in the fields not well covered by the other citation indexes. The results often seem to serve the function that is asked for. Harzing (2014) argues that GS is improving its database by significantly expanding the coverage, especially in chemistry, medicine, and physics. Other studies claim that GS can be used for performance analysis in the humanities and social sciences, if the data is professionally analyzed (Bornmann, Thor, Marx, & Schier, 2015; Prins *et al*. 2016). However, several studies have warned against misrepresentations and distortions when using GS for evaluation purposes (e.g., Delgado López-Cózar, Robinson-García, & Torres-Salinas, 2014; Jacso, 2012a; Harzing, 2012; Aguillo, 2011).

Delgado López-Cózar *et al.* (2014) showed that the results of GS can be manipulated: these authors "uploaded 6 documents to an institutional web domain that were authored by a fictitious researcher and referenced all publications of the members of the EC3 research group at the University of Granada. The detection of these papers by the crawler of Google Scholar caused an outburst in the number of citations included in the Google Scholar Citations profiles of the authors" (p. 446). GS has acknowledged the problem of being used as a data source for research evaluation purposes, and adds a warning underneath a scientist's profile: "Dates and citation counts are estimated and are determined automatically by a computer program."

The extent to which the lack of transparency of the underlying data sources of GS is a problem depends critically on the context of the evaluation at hand. An obvious advantage of GS from the perspective of managers and scientists (groups 3 and 4) is the inclusion of publications of all types and from all disciplines. The obvious source of error from the perspective of producers (group 1) and bibliometricians (group 2) is the lack of data cleaning and deduplication and the lack of transparency of the calculation of indicators.

*2.2. Disambiguation*

Although there are differences between the fee-based databases WoS and Scopus and the freely accessible product GS, there are also common problems. Records in these databases may be erroneous for a number of reasons. Citations can be missed because of misspellings by citing authors, or because of problems with the disambiguation of common author names (e.g., Jones, Singh, or Park). Those interested in a bibliometric study of individual scientists may be able to handle these problems using specific software, such as Publish & Perish for GS (Harzing, 2007).

Whereas the shortcomings may be well-known to producers (group 1) and bibliometricians (groups 2), they are frequently not sufficiently known among managers (group 3) and the scientists under study (group 4). Table 1, for example, lists 24 name variants in the address information of 141 documents retrieved from WoS with the search string "og= INST SCI TECH INFORMAT CHINA", 29 June 2016. Unlike the search tag "oo=", "og=" refers to institutional information indexed and consolidated by Thomson Reuters. Nevertheless, different institutions and different representations of the same institution are indicated.

**Table 1**. 24 name variants in the addresses among 141 records downloaded with the search string "og= INST SCI TECH INFORMAT CHINA" from WoS, 29 June 2016.

| |
|---|
| Inst Sci & Tech Informat China, Ctr Resource Sharing Promot, Beijing, Peoples R China |
| Inst Sci & Tech Informat China, Ctr Sci & Technol Studies, Beijing 100038, Peoples R China |
| Inst Sci & Tech Informat China, Informat Resource Ctr, Beijing, Peoples R China |
| Inst Sci & Tech Informat China, Informat Tech Supporting Ctr, Beijing 100038, Peoples R China |
| Inst Sci & Tech Informat China, Informat Technol Support Ctr, Beijing 100038, Peoples R China |
| Inst Sci & Tech Informat China, Informat Technol Supporting Ctr, Beijing 100038, Peoples R China |
| Inst Sci & Tech Informat China, Informat Technol Supporting Ctr, Beijing 700038, Peoples R China |
| Inst Sci & Tech Informat China, Informat Technol Supporting Ctr, Beijing, Peoples R China |
| Inst Sci & Tech Informat China, IT Support Ctr, Beijing 100038, Peoples R China |
| Inst Sci & Tech Informat China, IT Support Ctr, Beijing, Peoples R China |
| Inst Sci & Tech Informat China, Methodol Res Ctr Informat Sci, Beijing 100038, Peoples R China |
| Inst Sci & Tech Informat China, Res Ctr Informat Sci Methodol, Beijing 100038, Peoples R China |
| Inst Sci & Tech Informat China, Res Ctr Informat Sci Methodol, Beijing, Peoples R China |
| Inst Sci & Tech Informat China, Res Ctr Strateg Sci & Technol Issues, Beijing 100038, Peoples R China |
| Inst Sci & Tech Informat China, Res Ctr Strateg Sci & Technol Issues, Beijing, Peoples R China |
| Inst Sci & Tech Informat China, Strategy Res Ctr, Beijing 100038, Peoples R China |
| INST SCI & TECH INFORMAT CHINA,BEIJING 100038,PEOPLES R CHINA |
| INST SCI & TECH INFORMAT CHINA,BEIJING,PEOPLES R CHINA |
| INST SCI & TECH INFORMAT CHINA,CHONGQING BRANCH,CHONGQING,PEOPLES R CHINA |
| INST SCI & TECH INFORMAT CHINA,INT ONLINE INFORMAT RETRIEVAL SERV,BEIJING,PEOPLES R CHINA |
| INST SCI & TECH INFORMAT CHINA,PEKING,PEOPLES R CHINA |
| INST SCI & TECH INFORMAT CHINA,POB 3829,BEIJING 100038,PEOPLES R CHINA |
| INST SCI TECH INFORMAT CHINA,DIV INT RELAT & COOPERAT,BEIJING,PEOPLES R CHINA |
| ISTIC Thomson Reuters Joint Lab Scientometr Res, Beijing 100038, Peoples R China |

Among the thousands of authors with "Singh" as a family name, 45 are distinguished in WoS with the first initial "A." Twenty-one of these authors have more than ten publications, among which 9,266 publications for the "author" listed as "A. Singh". One cannot expect these publications to be sorted unambiguously to individual authors without substantial investments in name disambiguation. Scopus' Author Name Identifier disaggregates the set into 1,908 authors with an exact match for the search "Singh" as family name and "A." as first initial. The top-10 of these are shown in Figure 1. The same search—"A Singh"—provided "about 125,000 results" when using GS, obviously as a cumulative result of more than one thousand authors with this name.

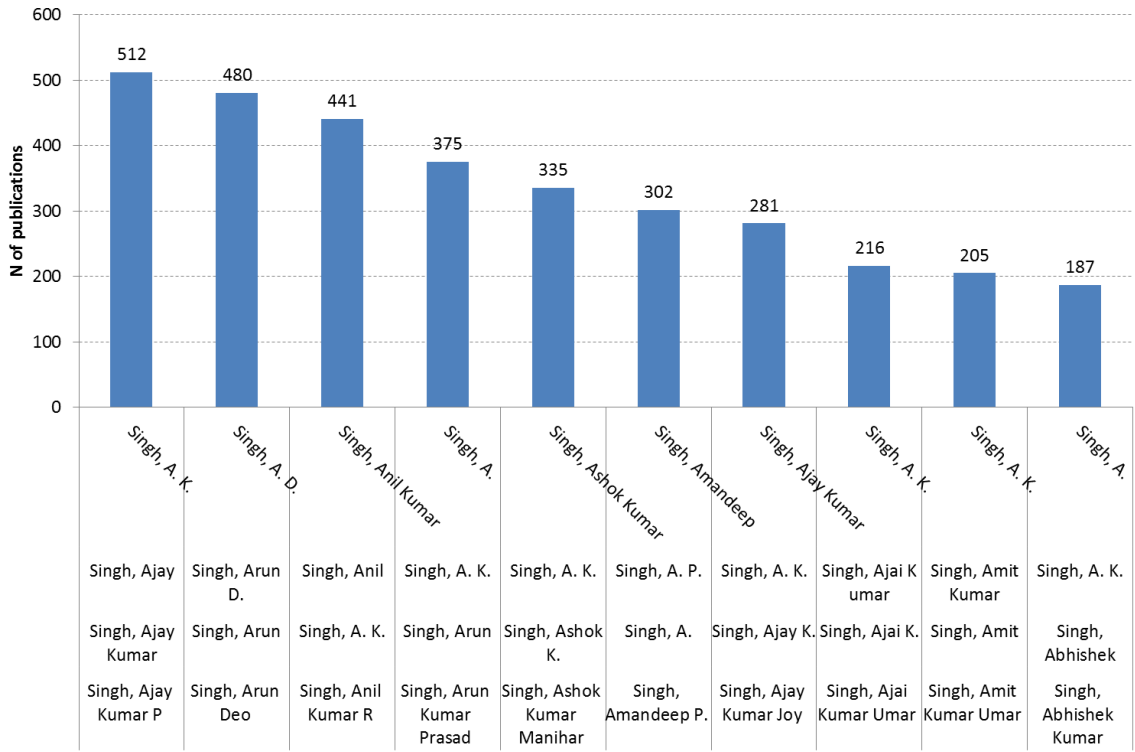

**Figure 1**. Numbers of documents of ten top authors with "Singh" as family name and "A" as first initial in Scopus (June 29, 2016). Main names attributed by the Scopus Author Identifier are added to the legend of the abscissa.

In order to clarify the resulting confusion, managers (group 3) are seeking and producers (group 1) are developing tools like ReseracherID (www.researcherid.com) and ORCID (http://orcid.org/). Thus, one hopes to be able soon to generate reliable and valid publication lists of single scientists and institutions. Bibliometricians (group 2) may also profit from these investments, because they can base their studies on data containing fewer errors from a technical point of view. Comparable initiatives to create unique identifiers for all digital objects in the databases may further decrease sources of error in the long run.

Whereas the problems of proper attribution of *publications* can thus perhaps be countered in local evaluation practices—by checking against the publication lists of active scientists (group 4)—uncertainty and possible error in the attribution of *citations* is even more systemic in citation analysis. Table 2 shows, for example, eleven journals in the Journal Citation Reports (JCR) of the Science Citation Index 2012 under names which are no longer valid.

Table 2. Not indexed journals and incorrect journal abbreviations with more than 10,000 citations in the JCR of the Science Citation Index 2012

| Journal | Citations | Reason why |
|---|---|---|
| *J Phys Chem-US* | 45,475 | no longer a single journal |
| *Lect Notes Comput Sc* | 42,723 | not a single journal |
| *Phys Rev* | 35,643 | no longer a single journal |
| *J Bone Joint Surg* | 28,812 | no longer a single journal |
| *Eur J Biochem* | 21,666 | renamed into *FEBS Journal* |
| *Biochim Biophys Acta* | 20,143 | no longer a single journal |
| *Method Mol Biol* | 17,870 | not covered |
| *P Soc Photo-Opt Inst* | 17,528 | not covered |
| *Mmwr-Morbid Mortal W* | 15,239 | not covered (weekly) |
| *J Biomed Mater Res* | 10,529 | no longer a single journal |
| *Communication* | 10,093 | no journal citation |

*Biochimica et Biophysica Acta*, for example, was first published in 1947, but since 1962 eight subtitles were derived alongside *Biochimica et Biophysica Acta-General Subjects.* The original title, however, is still cited 20,143 times in 2012; 532 times with publication year 2011, and 197 times with publication year 2010. These (523 + 197 =) 720 references are of the same order of magnitude as the references to these two years for the smallest among the subtitles (*Biochimica et Biophysica Acta-Reviews on Cancer*), which is cited (315 + 498) = 813 times from the two previous years. In other words, the noise in the citations of one journal can be of the same order as the signal of another journal.

In summary, both publication and citation data contain considerable technical error which is beyond control in evaluation studies. An evaluee who notes the omission or misplacement of one of her publications is not able to change the record within the time frame of an evaluation. An introduction to a special issue containing a review of the literature, for example, may be highly cited, but classified as an "editorial" and therefore not included in a professional evaluation (Leydesdorff & Opthof, 2011). The distinction between review and research articles in the WoS is based on citation statistics. While a name or affiliation can perhaps be corrected by evaluees, the assignments to document types or fields of science cannot be changed.[2]

Although these sources of error seem to be relevant to all four stakeholder groups, the implications for them are divergent. Whereas for producers (group 1) and bibliometricians (group 2) these problems can be considered as interesting puzzles that can be elaborated into research questions, the consequences for managers (group 3) are sometimes invisible, whereas for scientists (group 4) the consequences can be very consequential. Evaluees (group 4) may need to develop collaborations with bibliometricians (group 2) to unpack the reports and make the problems visible to management (group 3).

*2.3. Normalization*

Given that citation rates differ among fields of science and publication years, the normalization of citation impact has been studied since the mid-1980s (Schubert & Braun, 1986). While normalized indicators were initially used only by professional bibliometricians (group 2), they have recently been recommended for more general use (Hicks *et al.*, 2015) and have increasingly become the standard in evaluation practices (Waltman, 2016). For example, the World University Rankings of Times Higher Education (THE; at http://www.timeshighereducation.co.uk/world-university-rankings/) are based on normalized indicators, and these or other normalizations are also assumed in relatively new products, such as InCites (Thomson Reuters) and SciVal (Scopus) for use by managers (group 3) and scientists (group 4).

---

[2] WoS is consolidated each year with the appearance of the final version of the JCR in September/October of the following year.

Although field- and time-normalization of citation impact is nowadays considered a standard procedure among bibliometricians (group 2), there is little agreement about the details of the normalization procedures discussed. Different normalizations may lead to different outcomes (e.g., Ioannidis *et al.*, 2016). In the most common normalization procedure, for example, an expected citation rate is calculated for each paper by averaging the citation impact of all the papers that were published in the same subject category (journal set) and publication year. Two decisions have to be made: (1) which measure to use as the expected citation rate (e.g., the arithmetic mean), and (2) how to delineate the set of documents that will serve as the reference set. Since citation distributions are heavily skewed, one can build a strong case against the mean and in favor of non-parametric statistics (Bornmann & Mutz, 2011; Leydesdorff *et al.*, 2011; Seglen, 1992). Non-parametric statistics, however, are often less intuitively accessible for managers and evaluees.

Although it is possible to use the specific journal of the focal paper as the reference set, disciplines develop usually above the individual-journal level in terms of groups of journals. But how can one delineate a reference set at the above-journal level? Schubert and Braun (1986) proposed using the subject categories of WoS for this purpose (Moed, De Bruin, & Van Leeuwen, 1995).[3] However, these subject categories were developed decades ago for the purpose of information retrieval and evolved incrementally with the database (Pudovkin & Garfield, 2002). Using the WoS subject category "information science & library science," Leydesdorff and Bornmann (2016), for example, showed that these categories do not provide sufficient analytical clarity to carry bibliometric normalization in evaluations (cf. Van Eck, Waltman, van Raan, Klautz, & Peul, 2013).

The use of subject categories is stretched to its limit when reference values for specific sub-fields are needed (and a corresponding journal set does not exist) or in the case of multi-disciplinary fields and journals. Journals can be categorized in several fields; but at the paper level disciplinary affiliations may differ. Database providers (e.g., Thomson Reuters) try to solve this problem for certain multi-disciplinary journals, like *Nature* and *Science*: papers are assigned to subject categories on a paper-by-paper basis. However, this is not pursued for journals that are multi-disciplinary and subject-specific, such as *Angewandte Chemie – International Edition*. Subject categories from field-specific data bases, such as Chemical Abstracts (CA) for chemistry (Neuhaus & Daniel, 2009), PubMed for medicine (Rotolo & Leydesdorff, 2015), and Research Papers in Economics (RePEc) for economics (Bornmann & Wohlrabe, in preparation) have been proposed as an alternative. In these data bases, each paper is assigned to specific classes by experts in the discipline. The resulting classes can be used for alternative normalizations.

---

[3] The WoS subject categories were called ISI Subject Categories until the introduction of WoS v.5 in 2009.

Another basis for field delineations was proposed by Waltman and van Eck (2012b), who cluster the citation matrix among individual publications. The resulting clusters are used for the field-delineation in the current Leiden Ranking. This approach might improve the normalization when compared with journal-based schemes (Zitt *et al*., 2005). However, the latter are most commonly used in professional evaluations. For working scientists and managers (groups 3 and 4) this seems justified because of the role of the scientific journal as an archive, communication channel, gatekeeper and codifier of the field. As a result, the problem of the delineation of reference sets, though theoretically unsolved, is pragmatically handled in most applications of evaluative bibliometrics.

The network of aggregated citations is both horizontally differentiated in fields of science and hierarchically organized in strata such as elite structures (e.g., top-1%, top-10%, etc.). Such networks can be characterized as a complex system: whereas strongly connected nodes can be distinguished as groups, these groups are also connected (Simon, 1973; 2002). The decomposition of these systems are partially dependent on the theoretical framework of the analyst. This results in uncertainty in specific decompositions that can perhaps be specified (Rafols & Leydesdorff, 2009; Ruiz-Castillo & Waltman, 2015). Whereas the evaluations can be enriched and made more precise by investing in thesauri and indexes for names, addresses, and journal names, one can nevertheless always expect that managers (group 3) or evaluees (group 4) may rightfully complain about the, from their perspective, misclassification of specific papers.

Interdisciplinary efforts at the institutional or individual levels may suffer in particular from misclassifications in terms of disciplinary schemes at an aggregated level (Rafols, Leydesdorff, O'Hare, Nightingale, & Stirling, 2012). It should also be noted that citation normalization does not normalize for institutional differences between fields, such as the varying amounts of labour needed for a publication, the varying probability of being published in the top journals, etc. Although groups 1 and 2 may be aware of these limitations, users in groups 3 and 4 may have the mistaken impression that careful normalization solves all problems resulting from the incomparability of publication and epistemic cultures across fields of science.

In summary, the users of normalized citation scores should be informed that (1) different normalization procedures exist; (2) the procedures are controversial in discussions among bibliometricians (group 2); (3) different procedures lead to different results; and (4) an unambiguously "correct" solution for the decomposition of the database in terms of fields and specialties may not exist (Zitt, Ramanana-Rahary, & Bassecoulard, 2005). There is growing agreement about using non-parametric statistics in citation analysis, but with the possible exception of the percentage of publications in the top-10%, percentile-based measures are not yet commonly used.

*2.4. Citation windows*

In evaluations of individual scientists or institutions, the publications from the most recent years are of special interest to management (group 3). How did a given scientist perform during these most recent years? Furthermore, managers will be interested in the expected performance in the near future. These needs of institutional management do not correspond to the character of citation impact measurements. The urge for results from the most recent years and the technical standard in evaluative bibliometrics to use a longer citation window (Wang, 2013) can lead to tensions between managers (group 3) and their bibliometric staff or contractors (group 2).

A focus on the short-term citation (e.g., two or three years), means that contributions at the research front are appreciated more than longer-term contributions. Using Group-Based Trajectory Modeling (Nagin, 2005), Baumgartner and Leydesdorff (2014) showed that a considerable percentage of papers have mid-term and long-term citation rates higher than short-term ones. The authors recommend distinguishing between a focus on the short-term impact at the research front ("transient knowledge claims") and longer-term contributions ("sticky knowledge claims"). Empirical papers at a research front have to be positioned in relation to the papers of competing teams. These citations show the relevance of the knowledge claim in the *citing* paper, whereas citation impact is longer-term and associated with the relevance of the *cited* papers (Leydesdorff *et al*., 2016). However, the Journal Impact Factor (JIF) was explicitly defined by Garfield in terms of citations to the last two previous years of journal publications because in biochemistry the research front develops so rapidly (Bensman, 2007; de Solla Price, 1970; Garfield, 1972; Martyn & Gilchrist, 1968).[4] Journals and scientists that aim at higher scores on JIFs deliberately speed up the production cycle (e.g., *PLoS ONE*).

---

[4] JIFs are calculated by the scientific division of Thomson Reuters and are published annually in the JCR. To establish the JIF, the publications of a journal within a period of two years are taken into consideration and their citations are determined over the following year. The number of citations is then divided by the number of citable items.

It has become a common practice among bibliometricians (groups 1 and 2) to use a citation window of three to five years (Council of Canadian Academies, 2012). However, this window remains a pragmatic compromise between short- and long-term citation. Management (group 3) is interested not only in the activities at the research front, but also and especially in the long-term impact of a set of publications (by an institution or a scientist). However, long-term impact can only be measured in the long run. Using a citation window of three to five years, furthermore, implies that publications from the most recent three to five years—which are most interesting from the perspective of institutional management—cannot be included. This may be a problem for bibliometricians working for research organizations when confronted with pressure to deliver bibliometric analyses for recent years. An additional question is how the citation windows used in evaluation studies relate to the timeframe of knowledge creation in the subfield at hand. This may influence the extent to which these windows seem justified from the perspective of particularly group 4.

## 3. Ambivalences around the indicators currently in use

Whereas in the above our focus was on problems with the data and ambivalences about possible delineations—both over time (citation windows) and at each moment of time (e.g., fields of science), we now turn to problems about the various indicators that are commonly in use in evaluation practices.

*3.1. Journal Impact Factor*

Before the introduction of the *h*-index in 2005 (Hirsch, 2005)—to be discussed in a next section—metrics as a tool for research management had become almost identical with the use of JIFs, particularly in the biomedical and some social sciences. Although the JIF was not designed to evaluate the impact of papers or individuals, many institutional reports add JIFs to their list of papers.

> **Refereed articles (ISI)**
>
> Azrout, R., Van Spanje, J. H. P., & De Vreese, C. H. (2013). A threat called Turkey: Perceived religious threats and support for EU entry of Croatia, Switzerland and Turkey. *Acta Politica, 48*, 2-21.
> ► SSCI IMPACT FACTOR 0.361
>
> Azrout, R., Van Spanje, J. H. P., & De Vreese, C. H. (2013). Focusing on differences? Contextual conditions and anti-immigrant attitudes' effects on support for Turkey's EU membership. *International Journal of Public Opinion Research, 25*, 480-501.
> ► SSCI IMPACT FACTOR 1.125
>
> Beentjes, J. W. J., & Konig, R. P. (2013). Does exposure to music videos predict adolescents' sexual attitudes? *European Societies, 9*, 1-20.
> ► SSCI IMPACT FACTOR 0.548
>
> Bleakley, A., Piotrowski, J., Hennessy, M., & Jordan, A. B. (2013). Predictors of parents' intention to limit children's television viewing. *Journal of Public Health, 35*, 525-532.
> ► SSCI IMPACT FACTOR 1.993
>
> Bornmann, L., & Leydesdorff, L. (2013). Macro-indicators of citation impacts of six prolific countries: InCites data and the statistical significance of trends. *PLoS One, 8*, e56768.
> ► SCI IMPACT FACTOR 3.730
>
> Bornmann, L., & Leydesdorff, L. (2013). The validation of (advanced) bibliometric indicators through peer assessments: A comparative study using data from InCites and F1000. *Journal of Informetrics, 7*, 286-291.
> ► SSCI IMPACT FACTOR 4.153
>
> Bornmann, L., Leydesdorff, L., & Mutz, R. (2013). The use of percentiles and percentile rank classes in the analysis of bibliometric data: Opportunities and limits. *Journal of Informetrics, 7*, 158-165.
> ► SSCI IMPACT FACTOR 4.153
>
> Bornmann, L., Leydesdorff, L., & Wang, J. (2013). Which percentile-based approach should be preferred for calculating normalized citation impact values? An empirical comparison of five approaches including a newly developed one (P100). *Journal of Informetrics, 7*, 933-944.
> ► SSCI IMPACT FACTOR 4.153

**Figure 2**. Top of p. 83 of the Annual Report 2013 of the Amsterdam School of Communication Research (ASCoR) listing the refereed papers with their JIFs.

Figure 2, for example, cites the Annual Report 2013 of the institute to which one of us belongs. The refereed papers are carefully listed with the JIF for the previous year (2012) attributed to each of the publications. In the meantime (as of July 1, 2016), the first article in the left column of Figure 2 (Azrout, van Spanje, & de Vreese, 2013) has been cited three times, and the first article in the right column (Bornmann & Leydesdorff, 2013) ten times. The JIFs of the two

journals—*Acta Politica* and *PLoS ONE*, respectively—imply an expected citation impact of the latter paper ten times higher than that of the former. Thus, the measurement makes a difference.

Although the quality and influence of journals can be considered as a co-variate in the prediction of the *long-term* impact of papers, the JIF is nothing more than the average impact of the journal as a whole using the last two years of data. The expected value of the short-term impact as expressed by the JIF of a journal does not predict either the short- or the long-term impact of a paper published in it (Wang, 2013). JIFs may correlate poorly with the actual citation rates of most of the papers published in a journal (Seglen, 1997) since a few highly cited papers in a journal may have a very strong influence on the JIF. A study by Oswald (2007) of six economics journals shows that "the best article in an issue of a good to medium-quality journal routinely goes on to have much more citation impact than a 'poor' article published in an issue of a more prestigious journal" (p. 22). For this reason, some journals have decided to publish the full journal citation distribution rather than only the JIFs (Larivière *et al*., 2016).

The use of the JIF for the evaluation of individual papers provides an example of the so-called "ecological fallacy" (Kreft & de Leeuw, 1988; Robinson, 1950): inferences about the nature of single records (here: papers) are deduced from inferences about the group to which these records belong (here: the journals where the papers were published). However, an individual child can be weak in math in a school class which is the best in a school district. Citizen bibliometricians (in groups 3 and 4) may nevertheless wish to continue to use the JIF in research evaluations for pragmatic reasons, but this practice is ill-advised from the technical perspective of professional bibliometrics (group 2). In professional contexts, the JIF has been naturalized as a symbol of reputation of the journal, both by publishers and editors who advertise high values of the JIF as evidence of the journal's quality, and by researchers who may rightly see the fact that they are able to publish in a high-impact journal as a performance and recognition of their research prowess. In this sense, the JIF has perhaps become the boundary object par excellence.

Unlike WoS, Scopus uses the SNIP indicator as an alternative to JIF. "SNIP" stands for "Source-Normalized Impact per Paper"; but SNIP is a journal indicator and not an indicator of individual papers (Waltman, van Eck, van Leeuwen, & Visser, 2013). "Source-based citation analysis" refers technically to fractional counting of the citation in terms of the citing papers; that is, the sources of the citation. While the SNIP indicator is thus normalized at the level of papers, it is like the JIF an indicator of journals. Hence, unreflexive usage of the journal indicator SNIP for single-paper evaluations implies an ecological fallacy as serious as that affecting the use of JIFs on this level.

SNIP is a complex indicator that cannot be replicated without access to its production environment. The indicator was originally based on dividing a mean in the numerator ("raw impact per paper") by a ratio of the mean of the number of references in the set of citing papers (the so-called "database citation potential"; Garfield, 1979b) divided by the median of that set (Moed, 2010). In the revised version of SNIP (SNIP-2; Waltman *et al.*, 2013), the denominator is set equal to the harmonic mean of the numbers of cited references divided by three. Mingers (2014) noted that the choice of a harmonic mean instead of an arithmetic one (in the denominator, but not in the numerator) may strongly affect the resulting values.

The construction of new and allegedly more sophisticated indicators on the basis of recursive algorithms (such as PageRank) or by combining parametric and non-parametric statistics in a single formula has become common. For example, JCR provides the quartiles of JIFs as an indicator; Bornmann and Marx (2014) advocated using the median of the JIFs as the "Normalized Journal Position" (NJP) and claimed that NJPs can be compared across fields of science by "gauging the ranking of a journal in a subject category … to which the journal is assigned" (p. 496). As the authors formulated: "Researcher 1 has the most publications ($n = 72$) in *Journal 14* with an NJP of 0.19; for Researcher 2 …." (p. 496). One thus would use also in this case a (composed) measure at the aggregate level for measuring research performance at the individual level so that individual scholars can be ranked.

The underlying problem is that publications and citations cannot be directly compared; one needs a model for this comparison, and a model can always be improved or at least made more sophisticated. The model can be formulated as a simple rule, as in the case of the *h*-index, or in terminology and formulas that cannot be reproduced outside the context of the production of the indicator (e.g., SJR2). The resulting indicator will predictably be useful in some areas of science more than in others. As the problems manifest themselves in evaluation practices, the suppliers propose refinements (for example, the *g*-index as an improvement to the *h*-index; Egghe, 2006) or abandon the indicator in favor of a new version which improves on the previous one more radically (e.g., Abramo & D'Angelo, 2016).

*3.2. The h-index and its derivates*

In 2005, the *h*-index was introduced for measuring the performance of single scientists (Hirsch, 2005). Today, the index is very popular among managers (group 3) and scientists (group 4); it measures publications and citations as a single number. However, bibliometricians have identified a number of weaknesses. For example, the arbitrary definition has been criticized (Waltman & van Eck, 2012a): instead of counting *h* papers with at least *h* citations each, one could equally count *h* papers with at least $h^2$ citations.

The identified weaknesses of the *h*-index have not diminished its popularity. On the contrary, the index has been extensively used in an expanding literature of research papers with the aim of proposing optimized *h*-index variants (Bornmann, Mutz, Hug, & Daniel, 2011). However, the identification of many weaknesses has not led to abandoning the index, and none of the possibly improved variants have gained as much acceptance as the original index. The critique of the *h* index is a good example of the way in which the research results of bibliometricians (group 2) do not easily diffuse to managers and scientists (groups 3 and 4). As is the case with university rankings, innovations have to be actively translated and recontextualized by institutional stakeholders in order to have real-world effects.

The *h*-index is not normalized across fields, publication years of papers, or the professional age of scientists. Therefore, only scientists working in the same field, publishing in a similar time period, and having the same professional age can be compared. Valuable alternatives to the *h*-index—from a bibliometric perspective—include proposals for variants that are calculated on the basis of field-, time-, and age-normalized citation scores. For example, the number of papers of a scientist that was among the 10% most frequently cited papers within the same field and publication year (Bornmann & Marx, 2014; Plomp, 1990; Tijssen, Visser, & van Leeuwen, 2002; Waltman *et al.*, 2012) can be divided by the number of years since the scientist published her first paper. This would result in a number that is field-, time-, and age-normalized and can be used for the comparison of scientists from different fields and with different academic ages. As discussed above, however, field delineations are uncertain.

*3.3. Advanced bibliometric indicators*

The *h*-index and total citation counts are transparent indicators that can be understood and applied by managers (group 3) and scientists (group 4) with access to a literature database (e.g., WoS) without too many problems. In contrast, advanced bibliometric indicators are always part of a model which includes assumptions that can be used for the shaping of professional practices and demarcating them from common-sense. With the "standard practice" of using field- and time-normalized citation scores in evaluative bibliometrics, the professionally active bibliometricians (group 2) distinguish themselves from citizen bibliometrics (groups 3 and 4). To calculate these scores for larger units in science (e.g., research groups or institutions), one needs specialized knowledge in bibliometrics, as well as in-house and/or edited bibliometric data (e.g., from InCites, Thomson Reuters, or SciVal, Elsevier). A noteworthy trend is the inclusion of these more advanced indicators in standard databases.

Bibliometricians (group 2) and producers (group 1) have invested considerable resources into making the indicators "robust" and therefore socially acceptable (Hicks *et al.*, 2015; cf. Latour, 1987). The current rankings and databases, however, tend to build an institutional momentum that may tilt the balance between groups 1 and 2. Griliches (1994) called this "the computer paradox": "the measurement problems have become worse despite the abundance of data" (p. 11). The availability of increasing amounts of ("big") data increases the complexity of the indicators debate.

Whereas the mean normalized citation score—the single most widely used impact indicator for bibliometricians (group 2)—can still be explained to managers (group 3) and scientists (group 4), this becomes difficult with sophisticated indicators, such as percentiles of citations (Bornmann, Leydesdorff, & Mutz, 2013) or "citing-side" indicators (Waltman & van Eck, 2013). The percentile of a focal publication is the percentage of publications being published in the same year and subject category (the reference set) with fewer citations. A transformation of the citation curves in terms of percentiles is first needed (Leydesdorff & Bornmann, 2011) which may not be intuitively clear to groups 3 and 4.

Citing-side indicator scores are even more difficult to understand and reproduce than percentiles. Furthermore, their calculation using WoS or Scopus data can be expensive. Normalizing on the citing side means the division of the citation of a publication under study by the numbers of cited references in the citing documents. The first variant used the number of references of the citing paper; more complicated variants use the number of *active* references, that is, cited references with a direct link to an item in the source journals of WoS or Scopus. Waltman and van Eck (2013) have shown that these complicated variants have very good properties in terms of field- and time-normalization. However, the normalized scores that result from these calculations may be elusive for users (groups 3 and 4) since the link to raw citation counts as the conventional impact measure is lost. The use of these advanced indicators makes it necessary to involve a professional bibliometrician (group 2) in each evaluation process.

In other words, one generates a dependency relation which may work asymmetrically in the relations to management and the evaluees. Alternatively, one may not be able to use the results because they cannot be explained effectively. As a result, management (group 3) and scientists (group 4) may then prefer not to depend on specialists for analyzing bibliometric data. The lack of transparency (group 2) and the industrial interests in the background (group 1) may easily lead to a lack of trust in the professional alternative to citizen bibliometrics using freely available data like GS.

In bibliometrics, the trust in experts (group 2) is undermined by the development of information systems by producers (group 1) and by the increased availability of web-based data and indicators as a side effect of generic information infrastructures. Thus, the tension between professional and citizen bibliometrics is continuously reproduced. The client-oriented systems lead to the impression among managers (group 3) and scientists (group 4) that everyone is equally qualified to analyze data professionally. However, these stakeholders are frequently unaware that the database one is using is also a "black box" comprising uncertainty in the data and unavoidable decisions in the models (see the discussion of data ambivalences above).

*3.4. Benchmarking scientists and institutions*

In addition to the need to use reference sets in citation analysis, indicator values for the units under evaluation (e.g., individual scientists or institutions) need to be benchmarked. The performance of individual scientists, for example, can only be measured meaningfully by comparing a given scientist with a reference group. Some attempts have been published about efforts to make this possible in a specific field of research (Coleman, Bolumole, & Frankel, 2012; El Emam, Arbuckle, Jonker, & Anderson, 2012). As noted, however, both author name disambiguation and field delineations are uncertain. Furthermore, reference groups can be delineated institutionally, in terms of scientific communities, or in terms of other contexts.

Thomson Reuters with ResearcherID and ORCID Inc. with ORCID ID (and others) are building up large-scale bibliometric data bases at the individual level. However, these databases cannot yet be used for sampling valid reference groups. For a valid reference group, it is necessary to have data for all the scientists within a specific field and time period. Since the author databases are filled in by the scientists themselves on a voluntary basis, the completeness of the reference group cannot be assumed.

The evaluation of institutions may be less affected by the problem of defining reference sets for the comparison. A large number of university rankings are available (including also non-university research institutions) which contain reference values. For example, the Leiden ranking publishes the proportion of papers for universities worldwide which belong to the 10% most frequently cited publications within a given subject category and publication year (see below). If one uses precisely the same method for calculating the 10% most frequently cited papers published by an institution under study (Ahlgren *et al.*, 2014), the position of the institution can be determined using the reference sets provided by the Leiden ranking. Similar numbers can be obtained from other providers (e.g., SciVal), but the results can be significantly different because of differences in the underlying databases or ranking methodologies.

*3.5. The (re-)construction of university rankings*

Global university rankings, such as the THE and Shanghai Ranking, have played an important role in reshaping higher education into a global competition for students, resources, and reputation (Hazelkorn 2011). Rankings can be considered as boundary objects that are used to translate a combination of qualitative and quantitative information into an ordered list in which not the value of the indicators but the relative position of the universities *vis-à-vis* each other becomes the dominant framework for their interpretation. This interpretative framework is quite robust with respect to the ambitions or goals of the producers of the rankings. In other words, although the rankings are produced by groups 1 and 2, the university managers (group 3) are strongly driven to discard most information that rankings contain and focus exclusively on the rank order.

This tends to obscure the fact that rankings are also based on model assumptions which may themselves be improved over time. For example, the field delineations may change and this has consequences for the normalization. For example, the Leiden Ranking 2013 was based on normalization using the (approximately 250) WoS subject categories. The Leiden Ranking 2014 was based on normalization across 828 clusters of direct citation relations, while in the Leiden Ranking 2015 this was refined to 3,822 "micro-fields" and further to 4,113 "micro-fields" in the Leiden Ranking 2016.

Despite these changes in the field definitions from year to year, the rankings for different years are highly and significantly correlated. Universities, however, are mainly interested in whether the ranks are improved or worsened, and thus in the differences between years. The differences may in individual cases be affected by the model changes more than the overall correlations for the set. To address this problem, the Leiden Rankings recalculate the historical development of the value of the indicators for each university based on the latest methodology.

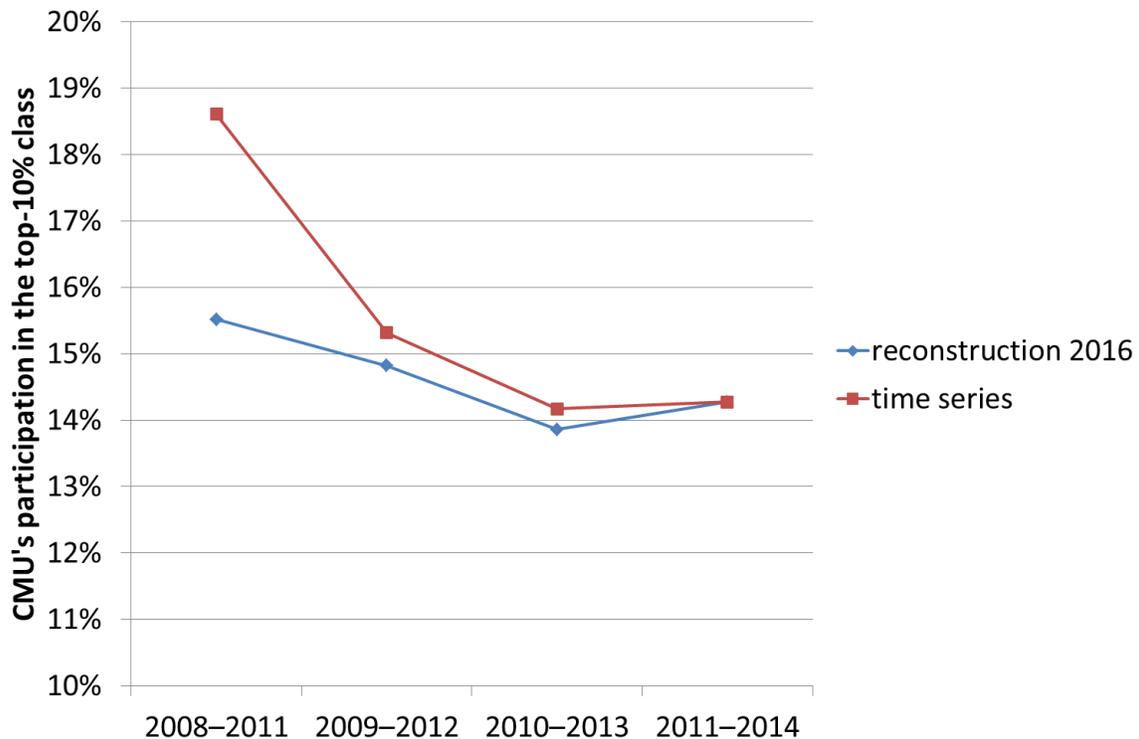

**Figure 3**: The participation of Carnegie Mellon University in the top-10% class of papers using the Leiden Rankings for subsequent years as a time series *versus* the reconstruction using the 2016-model; all journals included.

The largest change between 2013 and 2016 is found for Carnegie Mellon University (CMU). Figure 3 shows the different top-10% values of the CMU in the Leiden ranking published in the corresponding years (time series) and reconstructed in 2016 (reconstruction 2016). With 18.7% of its papers in the top-10% class (all journal included), this university was positioned on the 24$^{th}$ position in terms of this indicator in 2013 among 500 universities.[5] In 2016, the position of CMU has worsened to the 67$^{th}$ among 842 universities with a score of 14.3% of its papers in the top-10% class. The extension of the database has first a (trivial) effect on the ranks because universities may enter the domain with higher scores on the relevant parameter. Using the 2016-model, however, the value for 2013 is reconstructed as 15.5%, so that we can conclude that the difference is only 1.2% (that is, 15.5 – 14.3). The remaing 3.2% (that is, 18.7 – 15.5) of change is to be attributed to the change of the model. In other words, the model effect accounts for more than 70% of the change and the citation data themselves for 28%.

In summary, both the data and the model effects may cause differences between representations. As the results for the CMU show it would be a mistake to attribute changes over time only to increases or decreases in the citation scores themselves and not to the measurement instruments or underlying models. Whereas scientometric indicators serve the function of "objectivation" of the quality of discourse, "reification" is error-prone: differences may be due to changes in the data, the models, or the modeling effects on the data.

---

[5] On the website CMU is ranked at the 21$^{st}$ position in 2013 with the value of 17.3% for PP-top10%. The difference can be explained in terms of choosing all journals or only core journals as the domain.

The producers of the Leiden Rankings (group 1) actively discourage the users (group 3) to consider the rank numbers as the most important information; one is encouraged to inspect the underlying information about publications and citations. However, university managers (group 3) are keen on rankings to increase the standing of their institute in the global competition for reputation and resources. Hitherto, the dynamics in this process is still making university rankings more important whereas their relevance as single numbers can be questioned from the perspective of long-term knowledge production across different disciplines and specialties.

## 4. Conclusions

Different stakeholders are involved in bibliometric evaluations. The ambivalences around the use of data and indicators described in the sections above concern these stakeholders (we defined four groups) to variable extents. Errors and contingencies are possible and sometimes unavoidable when dealing with bibliometric data from each of these four perspectives.

(1) The first group concerns the **providers** of the data – mostly Thomson Reuters (WoS) or Elsevier (Scopus). The companies provide preprocessed data, but the provided data is not error-free. For example, former issues of journals are often missing and the volume, issue, and page numbers of many publications are frequently wrong (Marx, 2011). If advanced bibliometric indicators are used (e.g., field- and time-normalized citation scores), the institutions specialized in bibliometrics frequently provide these indicators which are continually produced by their in-house databases (based on WoS and/or Scopus data). However, users of the indicators should keep in mind that the calculation of advanced indicators is based on parameter choices that can be contingent and/or erroneous. For example, providers have to decide which field delineation method is applied (e.g., journal sets) and how.

(2) The second group are the **bibliometricians** who work at research institutions and analyze the data for evaluation purposes (e.g., in their role of providing advice for the selection of a candidate for a professorship). Bibliometricians can work differently with publication and citation data in many respects. They may not be aware of the problems or may have different standards. For example, they may include citation data in impact analyses for recent years or not. Some bibliometricians may have more acutely in mind that the underlying data are as a rule skewed, so that non-parametric statistics should be used instead of averages. Bibliometricians also use different field- and time-normalized indicators, which lead to different results.

(3, 4) The third and fourth groups are **managers** and **scientists** who use the bibliometric analyses, e.g., in an informed peer-review process or for a fellowship program to choose among different candidates. On the one hand, their bibliometric knowledge is often limited and there are many pitfalls in interpreting the bibliometric results. On the other hand, they are often far more knowledgeable about the research groups being evaluated and their disciplinary and institutional contexts; such knowledge is needed for the interpretation of the bibliometric results. Consequently, these two stakeholder groups are often more influential in evaluations than group 2. The difficulties entailed in the interpretation can perhaps be countered when the authors of the reports are able to describe the methods and results in a language that is less technical. However, this translation of specialist language into a more journalistic style can be considered as an art in itself.

In summary, bibliometric indicators are socially constructed in processes of translation among stakeholders. Two repertoires are then employed ( Gilbert & Mulkay, 1984): a constructivist one in which one "has to agree" upon the choice of indicators pragmatically, and a realist one considering indicators as windows on the world. Using this metaphor of a window, two tasks are specifiable: one has to improve the visibility through this window—in other words, the precision of the measurement—and one also has to keep visible the (re)construction of the window. When the construction is taken for granted, the evaluation becomes a technical problem that can eventually be black-boxed in a computer routine. Policy makers and research managers may like this perspective. However, such evaluations can be harmful for the processes under evaluation.

The irony of the situation is that an agreement with users sanctions the measurement results to such an extent that the evaluation reports sometimes no longer call attention to the problems by providing methodological reflections or indications of possible error. The indicators are then considered as reliable. The problems, however, are not only statistical, but also conceptual. For example, field delineations are uncertain and one can always dispute the choice of citation windows. The analyst makes choices which matter for the outcome of the evaluations. Whereas Hicks *et al.* (2015) argues in favor of consensus for certain basic principles in using bibliometric evaluations, we wish to note that a consensus can also lead to a compromise which hides the problems in the constructions. The indicators are constructs which offer a window on what is indicated. From the perspective of professional scientometrics, their main function is to indicate issues that require further investigation.

The measurement informs us about the status of a paper in the distribution. For example, it may be part of the 10% most frequently cited papers in a set. This formal result requires interpretation. Unlike the citizen bibliometrician, the professional bibliometrician has access and competence in a literature which can be used to carry an inference; for example, that the difference from a paper in the 89$^{th}$ percentile is significant. Without this discussion, the indicator is reified and possible choices are black-boxed. In our opinion, evaluation reports should provide openings for discussion among stakeholders by providing arguments and counter-arguments for the construction of a particular window among other possible ones (Stirling 2007). While bibliometric evaluations may be poor in registering past performance, they can thus serve to construct new perspectives.

**Acknowledgement**

We wish to thank Ludo Waltman for comments on a previous draft and interesting suggestions.